\documentclass{ifacconf}
\usepackage{amssymb}
\usepackage{amsmath}
\usepackage{subfigure}
\usepackage{subcaption}
\usepackage{graphicx}  
\usepackage[font=footnotesize]{caption}
\usepackage{natbib}

\begin{document}
\begin{frontmatter}

\title{Real-Time Solution-Seeking for Game-Theoretic Autonomous Driving via Time-Distributed Iterations} 
% Title, preferably not more than 10 words.

\thanks[footnoteinfo]{This work was supported by DARPA Young Faculty Award with the grant number D24AP00321.}

\author[]{Shaoqing Liu and Mushuang Liu} 

\address[]{Virginia Polytechnic Institute and State University, 
   Blacksburg, VA (e-mail:lshaoqing@vt.edu, mushuang@vt.edu).}

\begin{abstract}                % Abstract of 50--100 words
%This paper studies Newton and Newton--Kantorovich methods in time-distributed  game-theoretic model predictive control (GT-MPC) for autonomous driving. 
Computational complexity has been a major challenge in game-theoretic model predictive control (GT-MPC), as real-time solutions to a game (e.g., Nash equilibria (NEs)) have to be computed at each sampling instant of an MPC. This challenge is especially critical in autonomous driving, where interactions may involve many agents, and decisions must be made at fast sampling rates. We show that this challenge can be addressed through time-distributed solution-seeking iterations designed based on Newton and Newton--Kantorovich methods. Specifically, the autonomous vehicle (AV) decision-making problem is formulated as a  GT-MPC problem. To ensure solution attainability,   a potential game framework is adopted. Within this framework, both potential-function optimization and better response dynamics are used to seek the NE. To enable real-time implementation, Newton and Newton--Kantorovich methods are employed to solve the optimization problems arising in the NE-seeking algorithms, with their iterations distributed over time. Numerical experiments on an intersection-crossing scenario demonstrate that the proposed methods achieve effective real-time performance.
\end{abstract}


\end{frontmatter}
%===============================================================================

\section{Introduction}
%First paragraph: Game-theoretic autonomous driving has been emerging as a promising solution to characterizing agents' interactions during the autonomous vehicles' (AVs') decision-making...

Autonomous vehicles (AVs) must make real-time decisions in traffic environments while interacting with other traffic agents \citep{annurev:/content/journals/10.1146/annurev-control-060117-105157}. In a traffic environment, one vehicle's driving performance depends not only on its own actions but also on the actions of other agents, resulting in a coupled decision-making problem. To model such agents' interactions, game-theoretic methods have been employed \citep{7993050,Li2018ACCIntersection,9329208}. One widely-used solution concept in a game-theoretic problem is Nash equilibrium (NE) \citep{pnas.36.1.48}, representing a stable status such that no individual can improve its performance by unilaterally changing its action.  In recent years, special classes of games, e.g., potential games, have been exploited in autonomous driving to ensure the pure-strategy NE existence and NE-seeking algorithm convergence \citep{10101707, 10324408,10811754}.

%Autonomous vehicles (AVs) must make real-time decisions in traffic environments while accounting for interactions with surrounding vehicles. In many driving scenarios, however, the behavior of one vehicle affects the feasible action sets of others. This is particularly important in interactive settings such as merging, crossing, and other multi-vehicle conflict scenarios. Such mutual dependence motivates the use of game-theoretic models, which describe the decision-making process through coupled objectives \citep{7993050,Li2018ACCIntersection}. Recent studies have demonstrated that game-theoretic formulations can provide effective decision-making tools for autonomous driving ingeneral multi-vehicle settings \citep{9329208,10324408}.

%Second paragraph: One key challenge to solving the formulated game-theoretic problem is that the solution has to be solves at each sampling time, which could be impossible for agile or large scale systems. For single-agent MPC problem, the computtaional chalenge has been handled from various angles. For example,...

Despite the merits of game-theoretic models, their online solution seeking remains a major challenge. In a receding-horizon implementation, the decision-making problem must be solved repeatedly at every sampling instant, which is computationally expensive \citep{LIAOMCPHERSON2020108973}. For single-agent MPC, computationally efficient solvers have been developed, including time-distributed optimization \citep{ZANELLI2021109901}, and fast second-order Newton-type methods \citep{1583100}. These approaches reduce the computational cost by distributing the computation over time, such that at each sampling instant, only a limited number of iterations are performed to approximate the solution, as opposed to a sufficiently large number of iterations to solve the optimization completely. One widely used iterative algorithm within such time-distributed implementations is the Newton methods \citep{liu2024input}, which can solve the first-order necessary condition of an optimization problem efficiently. Furthermore, compared to Newton methods, Newton--Kantorovich methods \citep{BUTTS2016253,8554149} reuse the Jacobian information across iterations so that the Jacobian does not have to be updated at each iteration, thereby having the potential to further reduce the computational cost. However, despite the great potential, time-distributed game-theoretic model
predictive control (GT-MPC)  has not been well-studied. Its effectiveness in many real-world systems, like autonomous driving, remains an open question.

To fill the gap, this paper develops time-distributed GT-MPC solvers using both Newton and Newton--Kantorovich methods for autonomous driving. Specifically, to solve the NEs in a GT-MPC problem resulting from the AV decision-making, both potential function optimization and better response dynamics algorithms are employed. To solve the necessary conditions of the optimization problem, both Newton and Newton--Kantorovich methods are employed and are designed to distribute their iterations over time. With these developments, at each sampling instant, only a limited number of iterations are performed to approximate the NE at each instant. These algorithms are then tested in a 5-vehicle intersection-crossing scenario to evaluate and compare the algorithm performance.

\section{Problem Formulation}
In this section, the AV decision-making problem is formulated as a GT-MPC problem. Two NE-seeking algorithms are then presented.
%% Formulate the AV decision-making problem as a game-theoretic MPC problem
%% last paragraph: it's generally challenging tosolve the game problem at each sampling time, especially for highly nonlinear and agile systems. 
\subsection{AV decision-making as a GT-MPC problem}

Consider a system with $N$ autonomous vehicles, indexed by
$\mathcal{I}=\{1,2,\dots,N\}.$ 
The dynamics of vehicle $i$ are described as 
\begin{equation}
x_i(t+1)=f_i\bigl(x_i(t),u_i(t)\bigr),
\label{eq: vehicle dynamics}
\end{equation}
where $i\in\mathcal{I}$,
$x_i(t) \in \mathcal{X}_i$ and $u_i(t) \in \mathcal{U}_i$ are, respectively, the state and control input of vehicle $i$ at step $t$, $\mathcal{X}_i$ and $\mathcal{U}_i$ are the state space and control input space of agent $i$.

For each vehicle \(i\in\mathcal I\) and each sampling instant \(t\), define the cumulative cost functional as $V_i^t:\mathcal S_i\times \mathcal S_{-i}\to \mathbb R$, where \(\mathcal S_i\) is the feasible strategy set of vehicle \(i\), $\mathcal S_{-i}=\prod_{j\neq i}\mathcal S_j$ is the joint feasible strategy set of all other vehicles. We denote \(\mathcal S=\prod_{i\in\mathcal I}\mathcal S_i\) as the joint feasible strategy set of all vehicles. At time \(t\), vehicle \(i\) solves the finite-horizon optimization problem to seek its optimal control sequence (also called strategy):
\begin{small}
\begin{equation}\label{eq:GT_MPC}
\begin{aligned}
\mathbf{u}_i^*(t)
=
\mathop{\arg\min}\limits_{\mathbf{u}_i(t)\in\mathcal{S}_i}
\quad &
V_i^t\bigl(\mathbf{u}_i(t),\mathbf{u}_{-i}(t)\bigr) \\
=
\mathop{\arg\min}\limits_{\mathbf{u}_i(t)\in\mathcal{S}_i}
\quad &
\sum_{\tau=t}^{t+T-1}
\ell_i\bigl(x_i(\tau),u_i(\tau),x_{-i}(\tau),u_{-i}(\tau)\bigr),
\end{aligned}
\end{equation}    
\end{small}where \(\ell_i\) denotes the instantaneous cost at each time step, $\mathbf{u}_i(t)=\{u_i(t), u_i(t+1), \cdots , u_i(t+T-1)\}$ is the control sequence (strategy) of vehicle $i$ over the prediction horizon $T$, $\mathbf u_{-i}(t)
$ is the control sequences of all vehicles except for $i$. At each $t$, agent $i$ solves its optimal control sequence $\mathbf{u}^*_i(t)$ according to \eqref{eq:GT_MPC}, implements the first action $u^*_i(t)$, and repeats the decision-making \eqref{eq:GT_MPC} with a shifted horizon at time $t+1$.

The coupled multi-agent optimization problem \eqref{eq:GT_MPC} is essentially a game-theoretic problem \citep{10101707}. We denote the game \eqref{eq:GT_MPC} as
\begin{equation}
\mathcal G_t
=
\left\langle
\mathcal I,\{\mathcal S_i\}_{i\in\mathcal I},\{V_i^t\}_{i\in\mathcal I}
\right\rangle,
\label{eq:game}
\end{equation}
and seek NEs to solve the game.

\begin{defn}[Nash equilibrium \citep{pnas.36.1.48}]\label{def:NE}
A joint \\strategy profile
\[
\mathbf u^*(t)=\bigl(\mathbf u_1^*(t),\mathbf u_2^*(t),\dots,\mathbf u_N^*(t)\bigr)
\]
is called a Nash equilibrium of the game \(\mathcal G_t\) if, for every \(i\in\mathcal I\) and every \(\mathbf u_i(t)\in\mathcal S_i\),
\begin{equation}
V_i^t\bigl(\mathbf u_i^*(t),\mathbf u_{-i}^*(t)\bigr)
\le
V_i^t\bigl(\mathbf u_i(t),\mathbf u_{-i}^*(t)\bigr).
\end{equation}
\end{defn}

%Accordingly, at each sampling instant \(t\), we seek a Nash equilibrium of the game \(\mathcal G_t\), denoted by $\mathbf u^*(t)$.The implemented control input is the first element of each equilibrium control sequence, denoted by \(u_i^*(t)\). The state is then updated by \eqref{eq: vehicle dynamics} and the same procedure is repeated at the next sampling instant.

\subsection{Nash equilibrium seeking algorithms}%% In general, such a game problem may not always permit a pure-NE. In paer [], a potential game framework was developed to overcome this challenge. According to Theorem 1 in [], if the reward design satisfies certain form: J=...+...
 %Then the resulting game is a PG and thus can be solved by potential game optimization algorithm, i.e., Algorithm \ref{Algorithm1}.
% If the game is not a potentil game, one may consider, e.g., better response dynamics, to solve the resulting game \eqref{1}. The better response algorithm is detailed in Algorithm \ref{Al2}.

%[Theorem 6 in \cite{my_potential}]

%However, as shown in \citep{10101707}, if the game is a potential game, then we can ensure the solution existence and algorithm convergence.
In general, a pure-strategy NE defined in Definition~\ref{def:NE} may not always exist, unless the game has some special structures, e.g., a potential game. According to Theorem~8 in \citep{10101707}, if the cumulative cost satisfies
\begin{small}
\begin{equation}
V_i^t\bigl(\mathbf{u}_i(t),\mathbf{u}_{-i}(t)\bigr)
=
\alpha V_i^{\mathrm{self}}\bigl(\mathbf{u}_i(t)\bigr)
+
\beta \sum_{j\ne i} V_{ij}\bigl(\mathbf{u}_i(t),\mathbf{u}_j(t)\bigr),
\label{eq:individual_cost}
\end{equation}    
\end{small}then the resulting game is a potential game. Here $V_i^{\mathrm{self}}$ depends only on $\mathbf{u}_i(t)$ and captures self-related driving objectives, such as tracking the desired speed or maintaining driving comfort. $V_{ij}\bigl(\mathbf{u}_i(t),\mathbf{u}_j(t)\bigr) = V_{ji}\bigl(\mathbf{u}_j(t),\mathbf{u}_i(t)\bigr)$ denotes the symmetric pairwise interactions and can be used to characterize, e.g., collision penalty.
The parameters \(\alpha,\beta\in\mathbb R\) are coefficients to balance the two terms.

The potential function associated with the individual cost \eqref{eq:individual_cost} is
\begin{small}
\begin{equation}
\Phi\bigl(\mathbf{u}(t)\bigr)
=
\alpha \sum_{i\in\mathcal I} V_i^{\mathrm{self}}\bigl(\mathbf{u}_i(t)\bigr)
+
\beta \sum_{i\in\mathcal I}\sum_{j<i} V_{ij}\bigl(\mathbf{u}_i(t),\mathbf{u}_j(t)\bigr).
\label{eq:potential_function}
\end{equation}    
\end{small}
By Lemma 3, 4 in \citep{10101707}, if a game is a potential game, then a pure-strategy NE always exists and the game can be solved by potential function optimization:
\begin{equation}
\mathbf{u}^*(t)
=
\mathop{\arg\min}\limits_{\mathbf{u}(t)\in\mathcal S}
\Phi\bigl(\mathbf{u}(t)\bigr).
\label{eq:potential_optimization}
\end{equation}
The  detailed  procedures of using potential function optimization to solve the game can be found in Algorithm~\ref{Algorithm1}.
\begin{algorithm}[h]
\noindent\rule{\linewidth}{1pt}
\vspace{-20pt}
\caption{Potential function optimization to solve the game \eqref{eq:GT_MPC}}
\vspace{-6pt}
\noindent\rule{\linewidth}{0.2pt}

\label{Algorithm1}
\textbf{Input:}

Agent set: $\mathcal{I}$;

Current system state: $x(t)$;

Prediction horizon: $T$;

\textbf{Output:}

A Nash equilibrium: $\mathbf{u}^*(t)$.

\textbf{Procedure:}

1: Construct potential function $\Phi\bigl(\mathbf{u}(t)\bigr)$ according to \eqref{eq:potential_function}.

2: Solve the optimization problem in \eqref{eq:potential_optimization} to obtain $\mathbf{u}^*(t)$.

%3: Apply the first control input of each optimal control sequence.

%4: Update the system state and repeat at the next sampling instant.
\vspace{-3pt}

\noindent\rule{\linewidth}{0.2pt}
\end{algorithm}
%%%%%%%%%%%%%%%%%%%%%%%%%%%%%%%%%%%%%%%%%%%%%%%%%%%%%%%
% we use better response in newton and nk, and due to the property of exact reference soln computed by GS, the reference is better response

If the game is not a potential game, one may use, e.g., better response dynamics to solve the game. Specifically, we define the better response set of vehicle \(i\) to $\mathbf{u}_{-i}(t)$ as 
\begin{small}
    \begin{equation}
\begin{aligned}
&\mathcal{B}_i
\bigl(\mathbf{u}_i(t), \mathbf{u}_{-i}(t)\bigr)\\&
=
\left\{
\widetilde{\mathbf{u}}_i(t)\in\mathcal{S}_i
\;\middle|\;
V_i^t\bigl(\widetilde{\mathbf{u}}_i(t),\mathbf{u}_{-i}(t)\bigr)
<
V_i^t\bigl(\mathbf{u}_i(t),\mathbf{u}_{-i}(t)\bigr)
-\epsilon_{\mathrm{cost}}
\right\},
\end{aligned}
\label{eq:br_mapping}
\end{equation}
\end{small}

\noindent
where $\epsilon_{\mathrm{cost}}$ is the cost improvement tolerance.
Accordingly, the joint better response set is defined by
\begin{equation}
\mathcal B\bigl(\mathbf u(t)\bigr)
=
\prod_{i\in\mathcal I}\mathcal B_i\bigl(\mathbf{u}_i(t),\mathbf u_{-i}(t)\bigr).
\label{eq:joint_br_mapping}
\end{equation}
A Nash equilibrium under this method is characterized by $\mathbf u^*(t)$.
The detailed procedures of using better response dynamics to solve the game \eqref{eq:GT_MPC} are presented in Algorithm~\ref{Algorithm2}. But due to the limited number of Newton--type iterations, we may not find the exact minimizer of each response subproblem.

\vspace{-8pt}
\begin{algorithm}[h]
\noindent\rule{\linewidth}{1pt}
\vspace{-20pt}
\caption{Better response dynamics to solve the game \eqref{eq:GT_MPC}}
\vspace{-6pt}
\noindent\rule{\linewidth}{0.2pt}

\label{Algorithm2}
\textbf{Input:}

Agent set: $\mathcal{I}$;

Current system state: $x(t)$;

Prediction horizon: $T$;

Maximum number of sweeps: $L_{\max}$.

Strategy change tolerance: $\varepsilon$;

\textbf{Output:}

A Nash equilibrium: $\mathbf{u}^*(t)$.

\textbf{Procedure:}

\noindent
1: Initialize $l=0$, $\mathbf{u}^{(0)}(t)$,
and $\mathrm{Stop}=\mathrm{False}$.\par

\noindent
2: \textbf{While}
$\mathrm{Stop}=\mathrm{False}$ and $l<L_{\max}$
\textbf{do}\par

\noindent
3. \hspace*{1em}\textbf{For} each $i \in \mathcal{I}$

% \noindent
% 4: \hspace*{2em}
% Compute
% $\widetilde{\mathbf{u}}_i(t)$
% using $\mathbf{u}_i^{(l)}(t)$ and fixed $\mathbf{u}_{-i}^{(l)}(t)$\par

\noindent
4: \hspace*{2em}
Seek $\widetilde{\mathbf{u}}_i(t)$ from $\mathop{\arg\min}\limits_{\mathbf{u}_i(t)\in\mathcal{S}_i}
V_i^t\!\left(\mathbf{u}_i(t),\mathbf{u}_{-i}^{(l)}(t)\right)$

\noindent
5: \hspace*{2em}
\textbf{If}
$\widetilde{\mathbf{u}}_i(t)\in
\mathcal{B}_i\bigl(
\mathbf{u}_i^{(l)}(t),
\mathbf{u}_{-i}^{(l)}(t)
\bigr)$ in \eqref{eq:br_mapping}\par
\noindent\hspace*{3em}\textbf{Then}
Update $\mathbf{u}_i^{(l+1)}(t)
$ using $ \widetilde{\mathbf{u}}_i(t)$.\par

\noindent\hspace*{3em}
\textbf{Else}\par

\noindent\hspace*{4em}
$\mathbf{u}_i^{(l+1)}(t)
\leftarrow\mathbf{u}_i^{(l)}(t)$.\par

\noindent\hspace*{3em}
\textbf{End if}\par

\noindent\hspace*{2em}
\textbf{End for}\par

\noindent
6 :\hspace*{1em}
\textbf{If}
\begin{equation}
\left\|
\mathbf{u}^{(l+1)}(t)-\mathbf{u}^{(l)}(t)
\right\|_{\infty}
<\varepsilon.
\label{eq:stopping}
\end{equation}
\noindent\hspace*{2em}
\textbf{Then} Set $\mathrm{Stop}=\mathrm{True}$.\par

\noindent\hspace*{2em}
\textbf{End if}\par

\noindent
7: \hspace*{1em}Set $l\leftarrow l+1$.\par

\noindent\hspace*{1em}
\textbf{End While}\par

\noindent
8: Return $\mathbf{u}^{*}(t)=\mathbf{u}^{(l)}(t)$.\par

\vspace{-3pt}
\noindent\rule{\linewidth}{0.2pt}

\end{algorithm}
\section{Newton-Type Methods for GT-MPC}
This section presents time-distributed Newton-type methods for solving the GT-MPC problem. %We first rewrite the first-order conditions of the considered problems as nonlinear systems of equations. Based on this formulation, we introduce the time-distributed Newton and Newton--Kantorovich methods.
%标题 Time-Distributed Newton-Type Methods for Game-Theoretic MPC
%% Introduce time-distributed solver idea.
%%% Introduce Newton Method
%%% Introduce Newton--Kantorovich Method
\subsection{Time-distributed solution-seeking}

At each sampling instant $t$, the GT-MPC problem requires the computation of a control sequence $\mathbf{u}^*(t)$ to solve the game $\mathcal G_t$. Similar to solving optimization problems, the NE-seeking could also be performed by iterative algorithms such as Newton-type methods. With a time-distributed solution-seeking, we first warmstart the iterative algorithms using the solution from $t-1$, and then perform a limited number of iterations to approximate the solution at $t$ (as opposed to a sufficiently large number of iterations to accurately solve the game $\mathcal G_t$). To be more specific, we first warmstart the iterations at $t$ as %let \(\mathbf{u}^{(0)}(t)\) denote the initial guess for the iterative solver, and $\mathbf{u}^{(k)}(t)$ denote the $k$-th iterate at time $t$.
\begin{equation}
\mathbf{u}^{(0)}(t)=\mathcal{P}\bigl(\mathbf{u}^{(K)}(t-1)\bigr),
\label{eq:td_shift}
\end{equation}
where \(\mathbf{u}^{(0)}(t)\) represents the initial point of the iterative solver at $t$,  $\mathcal{P}$ denotes the horizon-shifting operator, and $\mathbf{u}^{(K)}(t-1)$ is the approximate solution from $t-1$ (i.e., $K$ iterations were performed at $t-1$). After the warmstart, we then iterate  $\mathbf{u}(t)$ with:
\begin{equation}
\mathbf{u}^{(k+1)}(t)
=
\mathbf{u}^{(k)}(t)
+
\Delta \mathbf{u}^{(k)}(t),
\label{eq:td_update}
\end{equation}
where $\Delta \mathbf{u}^{(k)}(t)$ is the correction computed by the iterative solver. The iteration \eqref{eq:td_update} is repeated until $k=K$ to derive $\mathbf{u}^{(K)}(t)$ as the approximate solution to the game $\mathcal G_t$ in \eqref{eq:game}.

In this way, the computational effort is distributed over time, and the iterative algorithms can be designed based on, e.g., Newton and Newton--Kantorovich methods.

\subsection{Newton method for the iterative algorithm}
%Consider the optimization pronblems involves in either the potential function optimization algorithm (i.e,m, Alhorithm \ref{}) or better resonpse dynamics algorithm (i.e., Algorithm \ref{}). For each optimization problem (i.e., \eqref{} in Algorithm \ref{}, or \eqref{} in Algorithm \ref{}), we write its first-order stationary condition as a nonlinear sytem of equation \citep{liu2024input}:

Consider the optimization problems involved in either the potential function optimization algorithm (i.e., Algorithm~\ref{Algorithm1}) or the better response dynamics algorithm (i.e., Algorithm~\ref{Algorithm2}). For each optimization problem (i.e., \eqref{eq:potential_optimization} in Algorithm~\ref{Algorithm1}, or \eqref{eq:br_mapping} in Algorithm~\ref{Algorithm2}), we write its first-order stationarity condition as a nonlinear system of equations \citep{liu2024input}:
\begin{equation}
0\in \mathcal F\bigl(\xi(t)\bigr)+\mathcal{N}\bigl(\xi(t)\bigr),
\label{eq:newton_root}
\end{equation}
where \(\mathcal{F}\bigl(\xi(t)\bigr)\) is the gradient of the cost function in the optimization problem with respect to \(\xi(t)\), $\xi$ is the decision variable in the optimization problem (i.e., \(\xi(t)=\mathbf u(t)\) in the Algorithm \ref{Algorithm1}, and 
\(\xi(t)=\mathbf u_i(t)\) in Algorithm \ref{Algorithm2}), and  \(\mathcal{N}(\xi(t))\) denotes the normal cone mapping associated with the feasible set of \(\xi(t)\) \citep{liu2024input}.

Let \(\xi^{(k)}(t)\) denote the \(k\)-th iterate at the sampling instant $t$. Newton's method linearizes \(\mathcal F\) at \(\xi^{(k)}(t)\), which gives
\begin{equation}
\mathcal F\bigl(\xi^{(k)}(t)\bigr)
+
J_{\mathcal F}^{(k)}
\bigl(\xi(t)-\xi^{(k)}(t)\bigr)
+
\mathcal{N}\bigl(\xi(t)\bigr)
\ni 0,
\label{eq:newton_linearization}
\end{equation}
where
\begin{equation}
J_{\mathcal F}^{(k)}
=
\frac{\partial \mathcal F}{\partial \xi}
\bigl(\xi^{(k)}(t)\bigr),
\label{eq:newton_jacobian}
\end{equation}
is the Jacobian matrix of \(\mathcal F\) at \(\xi^{(k)}(t)\).

The solution to \eqref{eq:newton_linearization} gives $\xi^{(k+1)}(t)$, and thus leads to $\Delta \mathbf{u}^{(k)}(t)$ in \eqref{eq:td_update}:
\begin{equation}
    \Delta \mathbf{u}^{(k)}(t) = \xi^{(k+1)}(t) - \xi^{(k)}(t).
    \label{eq:iterate_u}
\end{equation}
%To improve numerical robustness, a regularized linear system is used in the implementation.
%The Newton direction is computed from
%\begin{equation}
%\bigl(J_{\mathcal F}^{(k)}+\mu^{(k)}I\bigr)\Delta \xi^{(k)}(t)=-\mathcal F\bigl(\xi^{(k)}(t)\bigr),\label{eq:newton_regularized_direction}\end{equation}where \(\mu^{(k)}\ge0\) is a regularization parameter introduced to avoid singularity of \(J_{\mathcal F}^{(k)}\).
Specifically, in the potential function optimization algorithm, \(\xi(t)=\mathbf u(t)\), and 
\begin{equation}
\mathcal F\bigl(\mathbf u(t)\bigr)
=
\nabla_{\mathbf{u}} \Phi\bigl(\mathbf u(t)\bigr),
\label{eq:newton_potential_mapping}
\end{equation}
and
\begin{equation}
J_{\mathcal F}^{(k)}
=
\nabla^2_{\mathbf{u}} \Phi\bigl(\mathbf u^{(k)}(t)\bigr).
\label{eq:newton_potential_hessian}
\end{equation}
Thus, the Jacobian matrix of \(\mathcal F\) coincides with the Hessian matrix of the potential function $\Phi$.

In the better response algorithm, \(\xi(t)=\mathbf u_i(t)\), and 
\begin{equation}
\mathcal F\bigl(\mathbf u_i(t)\bigr)
=
\nabla_{\mathbf u_i}
V_i^t\bigl(\mathbf u_i(t),\mathbf u_{-i}(t)\bigr),
\label{eq:newton_br_mapping}
\end{equation}
and
\begin{equation}
J_{\mathcal F}^{(k)}
=
\nabla_{\mathbf u_i}^2
V_i^t\bigl(\mathbf u_i^{(k)}(t),\mathbf u_{-i}(t)\bigr).
\label{eq:newton_br_hessian}
\end{equation}
\subsection{Newton--Kantorovich method for iterative algorithm}
A limitation of the standard Newton method is that the Jacobian matrix $J_{\mathcal F}^{(k)}$ must be computed at every iteration $k$, which could be computationally expensive. One way to address this challenge is to employ the Newton--Kantorovich method \citep{8554149}, where the Jacobian matrix is computed only once at the initial iteration of each $t$ using potential function algorithm, or once in a better response sweep for each agent, then the hessian is reused in the subsequent iterations. %In the time-distributed setting, only a small number of Newton--Kantorovich iterations is performed at each sampling instant.

Let \(\xi^{(0)}(t)\) denote the initial point at \(t\). Define
\begin{equation}
J_{\mathcal F}^{(0)}
=
\frac{\partial \mathcal F}{\partial \xi}
\bigl(\xi^{(0)}(t)\bigr).
\label{eq:nk_jacobian}
\end{equation}
Then, the Newton--Kantorovich method replaces \(J_{\mathcal F}^{(k)}\) in the Newton iteration \eqref{eq:newton_linearization} by a fixed Jacobian \(J_{\mathcal F}^{(0)}\), which gives
\begin{equation}
\mathcal F\bigl(\xi^{(k)}(t)\bigr)
+
J_{\mathcal F}^{(0)}
\bigl(\xi(t)-\xi^{(k)}(t)\bigr)
+
\mathcal{N}\bigl(\xi(t)\bigr)
\ni 0.
\label{eq:nk_linearization}
\end{equation}

In the potential function algorithm, \(\mathcal F\bigl(\mathbf u(t)\bigr)\) is defined in \eqref{eq:newton_potential_mapping}, and
\begin{equation}
J_{\mathcal F}^{(0)}
=
\nabla_{\mathbf u}^2 \Phi\bigl(\mathbf u^{(0)}(t)\bigr).
\label{eq:nk_potential_hessian}
\end{equation}
In the better response algorithm, \(\mathcal F\bigl(\mathbf u_i(t)\bigr)\) is defined in \eqref{eq:newton_br_mapping}, and
\begin{equation}
J_{\mathcal F}^{(0)}
=
\nabla_{\mathbf u_i}^2
V_i^t\bigl(\mathbf u_i^{(0)}(t),\mathbf u_{-i}(t)\bigr).
\label{eq:nk_br_hessian}
\end{equation}
%%%%%% projection

% 要单独写br和potential方法吗Time-distributed Newton-type updates can be embedded into different game-solving structures. In this work, we consider both a potential-based formulation and a better response formulation. The corresponding implementations will be examined in the following numerical experiments.

\section{Numerical Experiments}
We consider specific autonomous driving scenarios in this section to illustrate the developed algorithms and to evaluate their performance. 

%In numerical experiments, we evaluate the proposed time-distributed Newton and Newton--Kantorovich methods under the two algorithms: potential function and better response algorithms. The experiments are designed to evaluate their approximation accuracy and computational efficiency.
\subsection{Simulation setup}
\begin{figure}[h]
\begin{center}
\includegraphics[width=4.2cm]{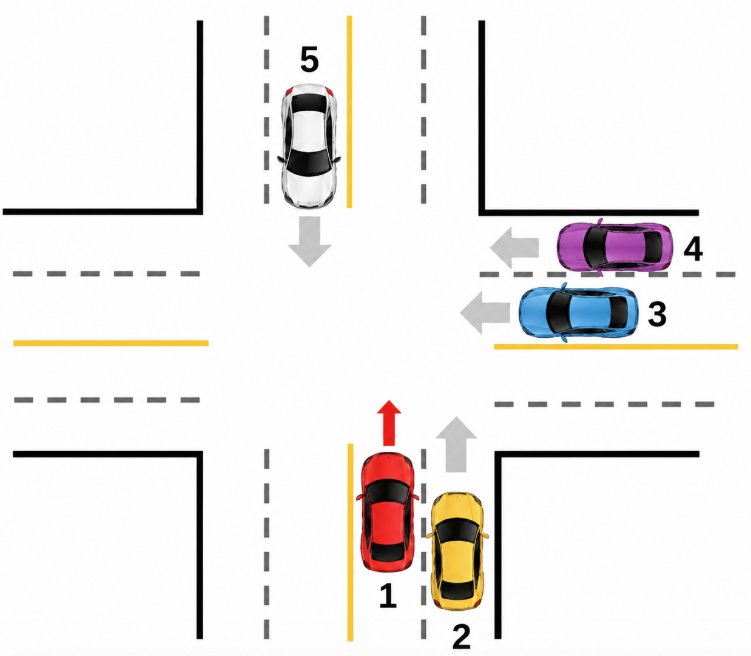}    % The printed column width is 8.4 cm.
\caption{Illustration of the intersection-crossing scenarios.}\label{fig:initial}
\end{center}
\end{figure}

We consider an intersection-crossing scenario involving $N=5$ vehicles, as shown in Fig.~\ref{fig:initial}.

The dynamics of each vehicle are modeled by a discrete-time linear function: 
\begin{equation}
x_i(t+1)=A x_i(t)+B u_i(t),
\label{eq:vehicle_dynamics}
\end{equation}
\[
A=
\begin{bmatrix}
1 & 0&\Delta t&0\\
0 & 1&0&\Delta t\\
0&0&1&0\\
0&0&0&1
\end{bmatrix},
\qquad
B=
\begin{bmatrix}
0&0\\
0&0\\
\Delta t&0\\
0&\Delta t
\end{bmatrix},
\label{eq:vehicle_dynamics_matrix}
\]

where \(x_i(t)=
\begin{bmatrix}
r_{i,x}(t)\\
r_{i,y}(t)\\
v_{i,x}(t)\\
v_{i,y}(t)
\end{bmatrix}\)
denote vehicle $i$'s positions and speeds along $x$ and $y$ axis, respectively, and \(u_i(t) = \begin{bmatrix}
    u_{i,x}\\
    u_{i,y}
\end{bmatrix}\) represents the acceleration.
We use $\left\|u_i(t)\right\|_2$ to represent the longitudinal acceleration along the vehicle's movement direction, and set $\|u_i(t)\|_2\le 5$,  for $\forall i \in\mathcal{I}$. %The initial speeds of the five vehicles are set to \(v(0)=[5.0,\ 4.5,\ 5.0,\ 4.5,\ 5.0]^\top\ \mathrm{m/s},\)where the \(i\)-th entry of \(v(0)\) is given by \(\left\|\begin{bmatrix}v_{i,x}(0)\\v_{i,y}(0)\end{bmatrix}\right\|_2\).
For the Newton and Newton--Kantorovich methods, the bound is ensured by projecting the elements of updated control sequence onto the interval after every iteration.

In our simulation, we consider the cost designed to be the form of \eqref{eq:individual_cost}. Specifically, the self-related cost, (i.e., $V_i^{\text{self}}$ in \eqref{eq:individual_cost}) is designed to track a desired speed:
\begin{equation}
V_i^{\mathrm{self}}\bigl(\mathbf{u}_i(t)\bigr)
=
\sum_{\tau=t}^{t+T-1}
\left(
\frac{v_i(\tau+1)-v_i^{\mathrm{ref}}}{v_i^{\mathrm{ref}}}
\right)^2.
\label{eq:self_cost}
\end{equation}
The interaction  term (i.e., $V_{ij}$ in \eqref{eq:individual_cost}) is designed to penalize small inter-vehicle distances:
\begin{equation}
V_{ij}\bigl(\mathbf{u}_i(t),\mathbf{u}_j(t)\bigr)
=
\sum_{\tau=t}^{t+T-1}
\frac{1}{\|r_i(\tau+1)-r_j(\tau+1)\|^2+\delta},
\label{eq:interaction_cost}
\end{equation}
where \(v_i^{\mathrm{ref}}\) is the reference speed of vehicle \(i\), and \(\delta>0\) is a small constant to avoid the denominator being zero.

At each sampling instant \(t\), we use the GlobalSearch solver in MATLAB \citep{mathworks_global_optimization_toolbox_2025b} to approximate the exact solution \(\mathbf{u}^{\mathrm{GS}}(t)\) and to do the comparison with the time-distributed solutions from Newton and Newton--Kantorovich methods. In the Newton-type methods, the control sequence at the first sampling instant $t=1$ is initialized by the GlobalSearch, while at later instants the initialization is obtained by warmstart (i.e., \eqref{eq:td_shift}). After the warmstart, a finite number of Newton-type iterations is performed to update the control sequence (i.e., \eqref{eq:td_update}).
Following the receding-horizon principle, only the first element in the control sequence (i.e., $u(t)$ in $\mathbf{u^*}(t)$) is applied to the system to move the system state to $x(t+1)$. 

The sampling period and the prediction horizon are selected as
$\Delta t=0.24~\mathrm{s}$ and $T=10$, respectively, corresponding to
a prediction duration of $T\Delta t=2.4~\mathrm{s}$. The relatively
short sampling period limits the state variation between two consecutive
sampling instants and thereby promotes the similarity between
the games $\mathcal{G}_{t-1}$ and $\mathcal{G}_{t}$. This
improves the effectiveness of the warmstart in \eqref{eq:td_shift}. %To ensure a fair comparison, the first instant is excluded from the evaluation. 
%The resulting approximate solutions at the remaining sampling instants are then compared with $\mathbf{u}^{\mathrm{GS}}(t)$.

% near iteartion part

\subsection{Potential function optimization to solve the game}
We consider the potential function optimization algorithm (i.e., Algorithm \ref{Algorithm1}) for the NE-seeking. We define the approximation errors of the control sequence from the Newton and Newton--Kantorovich methods, respectively, as 
\begin{equation}
e_{\mathrm{P},\mathrm{M}}^{(K)}(t)
=
\left\|
\mathbf{u}_{\mathrm{P},\mathrm{M}}^{(K)}(t)
-
\mathbf{u}_{\mathrm{P}}^{\mathrm{GS}}(t)
\right\|_2,
\label{eq:stepwise_error_potential_nk}
\end{equation}
where method $\mathrm{M}\in\{\mathrm{Newton},\mathrm{Newton-Kantorovich}\}$,  $\mathbf{u}_{\mathrm{P},\mathrm{M}}^{(K)}(t)$ denote the control sequences from the Newton (i.e., \eqref{eq:newton_linearization}) and Newton--Kantorovich (i.e., \eqref{eq:nk_linearization}) methods after $K$ iterations at each $t$, respectively, and $\mathbf{u}_{\mathrm{P}}^{\mathrm{GS}}(t)$ denotes the exact solution computed by GlobalSearch in the potential function optimization algorithm. 

\begin{figure}[h]
    \centering
    \begin{minipage}[t]{0.49\linewidth}
        \centering
        \includegraphics[height=3.1cm]{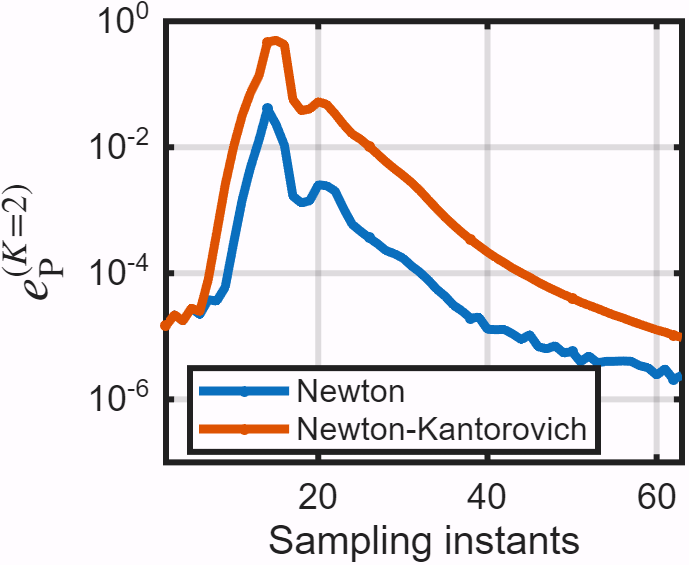}\\[-1pt]
        {\footnotesize (a)}
    \end{minipage}
    \hfill
    \begin{minipage}[t]{0.49\linewidth}
        \centering
        \includegraphics[height=3.1cm]{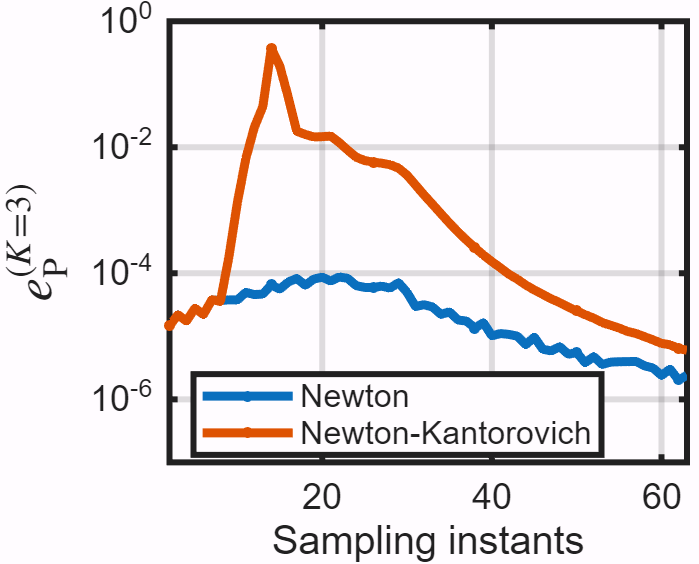}\\[-1pt]
        {\footnotesize (b)}
    \end{minipage}
    \caption{\footnotesize Approximation error at each $t$ using potential function optimization: (a) $e^{(K=2)}_\mathrm{P}$ and (b) $e^{(K=3)}_\mathrm{P}$.}
    \label{fig:potential_newton_nk_acc_time}
\end{figure}
Fig.~\ref{fig:potential_newton_nk_acc_time}(a) and Fig.~\ref{fig:potential_newton_nk_acc_time}(b) show the trajectories of the approximation error in \eqref{eq:stepwise_error_potential_nk}, when the iteration numbers are \(K=2\) and \(K=3\) respectively. The vertical axis is
shown on a logarithmic scale. Fig.~\ref{fig:potential_newton_nk_acc_time} leads to the following observations:

\begin{enumerate}
    \item For both figures, the approximation errors remain  small at the beginning ($t\leq 10$) and end ($t\geq 40$) of the trajectories; while larger errors are observed around $t\in(10,40)$, which corresponds to the time when the vehicles are around the middle of the intersection. This is because the games $\mathcal G_t$ at different $t$ have almost no change when the vehicles are far away from each other (i.e., $t\leq 10$ or $t\geq 40$) due to weak agents' interactions, but the changes become significant (from the order of $10^{-4}$ to $10^{0}$) when the vehicles are close (i.e., $t\in(10,40)$) due to strong interactions. This suggests that the time-distributed NE-seeking works well when the sequential games (i.e., $\mathcal G_t$ for sequential $t$) remain similar, while the errors increase if the games change significantly.
    \item Comparing the two methods, Newton generally produces smaller approximate errors than Newton--Kantorovich.  This difference is expected because Newton updates the Hessian approximation during each iteration, whereas Newton--Kantorovich uses the Hessian evaluated at the initial point. Therefore, when $\mathcal{G}_t$ changes significantly and the iterates move farther from the initial point, the fixed Hessian used by Newton--Kantorovich becomes less accurate.

\end{enumerate}

%relatively large approximately from $3$ s to $8$ s. This is consistent with the interaction cost in \eqref{eq:interaction_cost}: when conflicting vehicles are close to each other, the term $\frac{1}{(\|r_i(\tau)-r_j(\tau)\|^2+\delta)}$ becomes large, so the optimization problem is more sensitive to the control sequence and the error is larger. When the vehicles move away, the pairwise distances increase and the interaction cost decreases, so the error becomes small. For the Newton method, increasing the iteration number from $K=2$ to $K=5$ significantly reduces the error, and a similar trend is observed for the Newton--Kantorovich method. However, for the same iteration, the Newton--Kantorovich method yields larger errors than the Newton method, since it uses only the Jacobian evaluated at the first iteration and keeps it fixed in the subsequent updates.

\begin{figure}[h]
    \centering
    \includegraphics[width=0.65\linewidth]{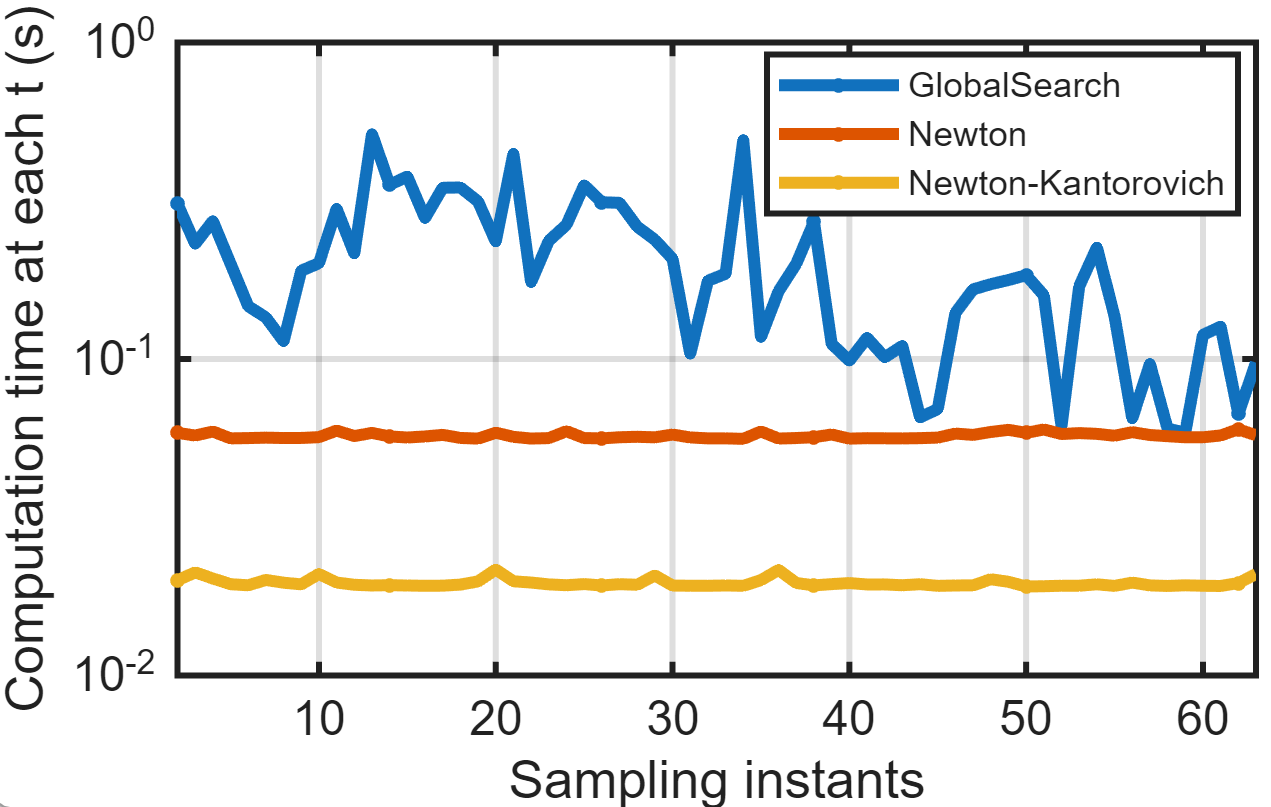}
    \caption{\footnotesize Computation time of GlobalSearch, Newton, and Newton--Kantorovich in potential function optimization algorithm.}
    \label{fig:potential_solver_time}
\end{figure}

To further quantify how the approximation accuracy varies with the number of iterations, we consider the mean error along the trajectory. For each fixed iteration $K$ and each method $\mathrm{M}\in\{\mathrm{Newton},\mathrm{Newton-Kantorovich}\}$, let $e_{\mathrm{P},\mathrm{M}}^{(K)}(t)$ denote the approximation error at $t$, then the mean error is defined as
\begin{equation}
\bar e_{\mathrm{P},\mathrm{M}}^{(K)}
=
\frac{1}{T_{\mathrm{sim}}-1}
\sum_{t=2}^{T_{\mathrm{sim}}}
e_{\mathrm{P},\mathrm{M}}^{(K)}(t),
\label{eq:mean_error_potential}
\end{equation}
where $T_{\mathrm{sim}}$ is the total number of sampling instants in the simulation trajectory.
%$\bar e_{\mathrm{P},\mathrm{M}}^{(K)}$ reflects the overall approximation across all evaluated sampling instants, while $e_{\mathrm{P},\mathrm{M},\max}^{(K)}$ characterizes the worst-case deviation from the exact solution. 
For Newton and Newton--Kantorovich methods, the mean error decrease as the iteration \(K\) increases. Compared with the Newton method, the Newton--Kantorovich method is less accurate due to the fixed Jacobian. 
% \begin{figure}
%     \centering
%     \includegraphics[width=0.5\linewidth]{potentialmeannew.png}
%     \caption{Enter Caption}
%     \label{fig:placeholder}
% \end{figure}
% \begin{figure}
%     \centering
%     \includegraphics[width=0.5\linewidth]{potentmeansequ.png}
%     \caption{Enter Caption}
%     \label{fig:placeholder}
% \end{figure}
\begin{figure}[h]
    \centering
    \includegraphics[width=0.63\linewidth]{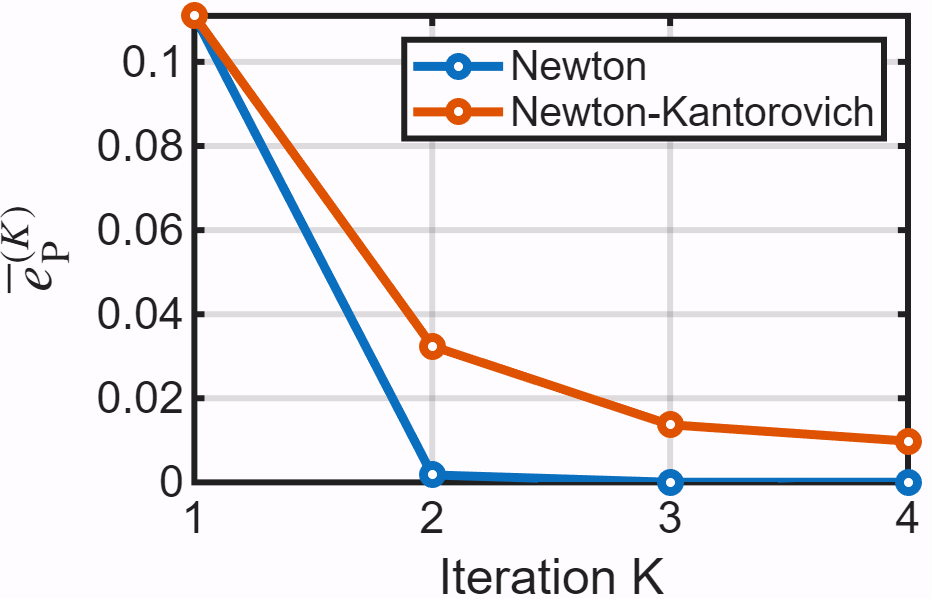}
    \caption{Mean error over the trajectory.}
    \label{fig:acceleration_error_convergence}
\end{figure}
 The computational cost of each iteration of the Newton--Kantorovich method is cheaper due to the fixed Jacobian. To verify this, we also plot the computation time of the three methods (GlobalSearch, Newton, and Newton--Kantorovich) in Fig.~\ref{fig:potential_solver_time}. 

% \begin{figure}
%     \centering
%     \includegraphics[width=0.5\linewidth]{potenttime.png}
%     \caption{Enter Caption}
%     \label{fig:placeholder}
% \end{figure}
 Note that for Newton and Newton--Kantorovich methods, we fix $K=3$. The computation time was recorded on a desktop computer equipped with an Intel Core Ultra 9 285K processor (3.70 GHz), 128 GB RAM, running MATLAB R2025b.
 As shown in Fig.~\ref{fig:potential_solver_time},  both the Newton and Newton--Kantorovich methods take much less computation time than GlobalSearch for the NE-seeking. Moreover, the Newton--Kantorovich method uses the least computation time to fulfill the iterations. %consistently faster than the Newton method by avoiding recomputing the Hessian matrix at every iteration.

%Under the potential function algorithm, this fixed Hessian approximation causes a larger error because the problem is solved in a centralized form. Even so, the Newton--Kantorovich method still shows clear improvement as the iteration increases.

\subsection{Better response dynamics to solve the game}
%We next evaluate the methods under the better response algorithm. For each \(i\in\mathcal I\), the self-related and interaction-related costs are defined as \eqref{eq:self_cost} and \eqref{eq:interaction_cost}.

We next consider the better response dynamics algorithm (i.e., Algorithm \ref{Algorithm2}) for NE-seeking. The approximation errors of the time-distributed solutions from Newton and Newton--Kantorovich methods are defined as
\begin{equation}
e_{\mathrm{BR},\mathrm{M}}^{(K)}(t)
=
\left\|
\mathbf{u}_{\mathrm{BR},\mathrm{M}}^{(K)}(t)
-
\mathbf{u}_{\mathrm{BR}}^{\mathrm{GS}}(t)
\right\|_2,
\label{eq:br_acc_error}
\end{equation}
where $\mathbf{u}_{\mathrm{BR},\mathrm{M}}^{(K)}(t)$  denotes the control sequence from \eqref{eq:newton_linearization} and \eqref{eq:nk_linearization} after $K$ Newton--type iterations using better response algorithm, and $\mathbf{u}_{\mathrm{BR}}^{\mathrm{GS}}(t)$ is the exact control sequence computed by GlobalSearch.
% \begin{figure}
%     \centering
%     \includegraphics[width=0.5\linewidth]{brerror2seq.png}
%     \caption{Enter Caption}
%     \label{fig:placeholder}
% \end{figure}
\begin{figure}[h]
    \centering
            \begin{minipage}[t]{0.49\linewidth}
        \centering
        \includegraphics[height=3.15cm]{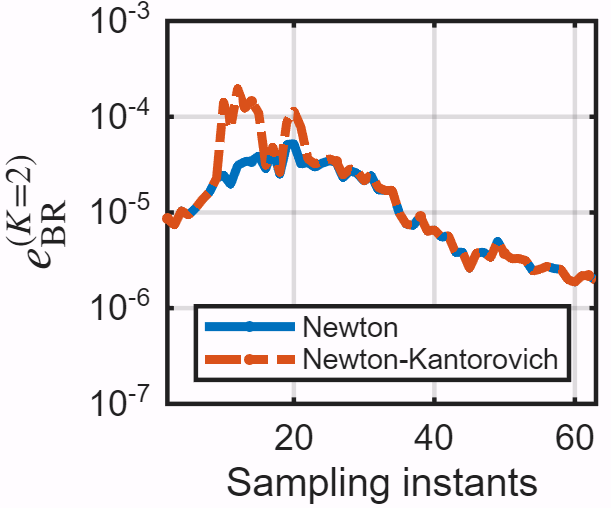}\\[-1pt]
        {\footnotesize (a)}
        \label{fig:brnewton}
    \end{minipage}
    \hfill
        \begin{minipage}[t]{0.49\linewidth}
        \centering
        \includegraphics[height=3.15cm]{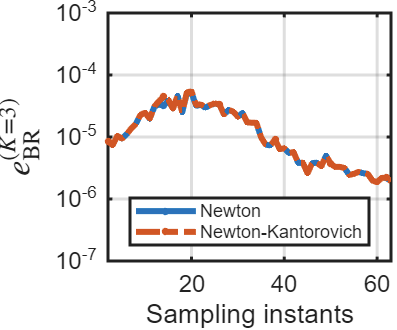}\\[-1pt]
        {\footnotesize (b)}
    \end{minipage}
    \caption{\footnotesize Approximation errors at each $t$ for different $K$ using better response algorithm: (a) $e_{\mathrm{BR}}^{(K=2)}$ and (b) $e_{\mathrm{BR}}^{(K=3)}$.}
    \label{fig:br_acc_error_time}
\end{figure}
% \begin{figure}
%     \centering
%     \includegraphics[width=0.5\linewidth]{br3seq.png}
%     \caption{Enter Caption}
%     \label{fig:placeholder}
% \end{figure}
Fig.~\ref{fig:br_acc_error_time}(a) and
Fig.~\ref{fig:br_acc_error_time}(b) show the trajectories of the
control sequence approximation error under the better response
algorithm for $K=2$ and $K=3$, respectively. The vertical axis is
shown on a logarithmic scale. Fig.~\ref{fig:br_acc_error_time} leads to the following observations:
\begin{enumerate}

    \item
    For both values of $K$, the approximation errors remain small at
    the beginning ($t\leq 10$) and near the end ($t\geq 40$), large errors appear over
    $t\in(10, 40)$.

    \item
    For both $K=2, 3$, the Newton and Newton--Kantorovich curves are
    close over most of the closed-loop trajectory, the Newton method generally gives a slightly smaller error. And from $K=2$ to $K=3$, the curves get even closer.
\end{enumerate}

    Under the better response algorithm, each vehicle updates only its
    own control sequence while the control sequences of the other
    vehicles are fixed. Therefore, each better response update involves
    a lower-dimensional optimization problem, rather than
    the optimization problem used in potential-function
    optimization. Consequently, the
    Hessian varies only moderately during an Newton type update, and
    the fixed Hessian used by Newton--Kantorovich remains close to the
    Hessian recomputed by Newton. This explains why the two methods
    generate similar control sequences under the better response
    algorithm.

%We first show how the error evolves for different iterations in Fig.~\ref{fig:br_acc_error_time}. For both Newton and Newton--Kantorovich under the better response algorithm, increasing the iterations leads to a clear reduction of the error. Moreover, compared with the potential function algorithm, the errors under the better response algorithm are much smaller overall. This is because the potential function algorithm solves a centralized problem, whereas the better response algorithm updates only a local subproblem with the other agents' control sequences fixed.

To evaluate how the approximation error varies with \(K\), we also plot the error statistics based on \eqref{eq:br_acc_error}. We define the mean error over the trajectory as
\begin{equation}
\bar e_{\mathrm{BR},\mathrm{M}}^{(K)}
=
\frac{1}{T_{\mathrm{sim}}-1}
\sum_{t=2}^{T_{\mathrm{sim}}}
e_{\mathrm{BR},\mathrm{M}}^{(K)}(t),
\label{eq:mean_error_br}
\end{equation}
where $e_{\mathrm{BR},\mathrm{M}}^{(K)}(t)$ denote the approximation error for the iterative method $M$. 
\vspace{-6pt}
\begin{figure}
    \centering
    \includegraphics[width=0.65\linewidth]{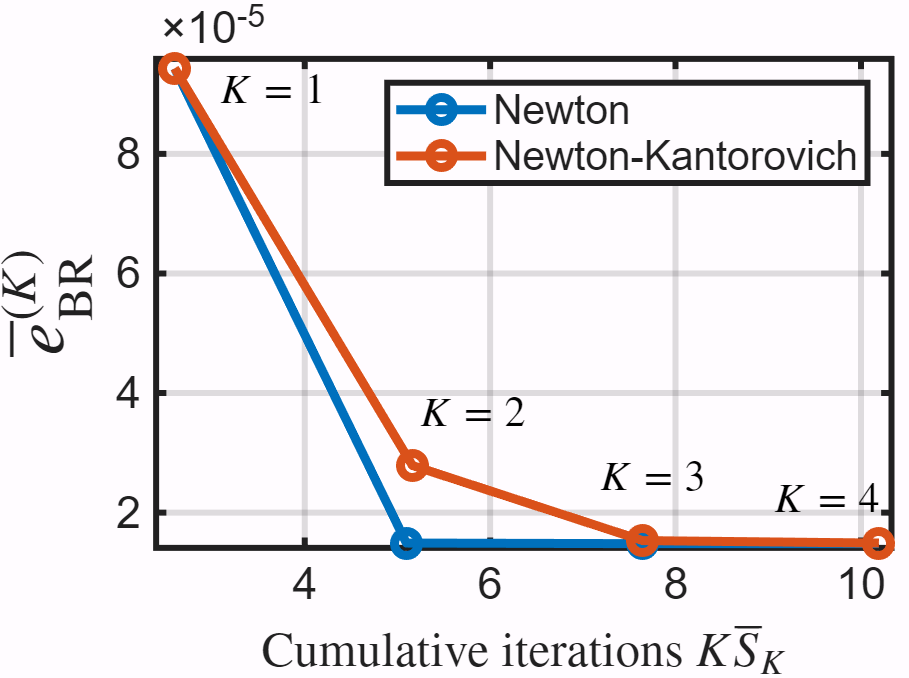}
    \caption{\footnotesize Mean error along the trajectory}
    \label{fig:acc_error_iter}
\end{figure}

% \begin{figure}
%     \centering
%     \includegraphics[width=0.5\linewidth]{brmeanseq.png}
%     \caption{Enter Caption}
%     \label{fig:placeholder}
% \end{figure}

As shown in Fig. ~\ref{fig:acc_error_iter}, mean sequence error decreases from
$K=1$ to $K=2$ and keeps numerically stable for Newton for $K=2,3,4$, Newton--Kantorovich needs more iterations than Newton to reach the same error, showing the methods' work under better response algorithm. $S_K(t)$ denote the number of better response sweeps performed at each $t$ for each agent, $\overline{S}_K
=
\frac{1}{T_{\mathrm{sim}}-1}\sum_{t=2}^{T_{\mathrm{sim}}}S_K(t).$ And $\overline{S}_K$ is approximately $2.5968$ for both methods at $K=1$. For $K=2$, it is approximately $2.5484$ for Newton and $2.5806$ for Newton--Kantorovich. For $K=3,4$, $\overline{S}_K$ is $2.5484$ for both methods.

\begin{figure}[h]
    \centering
            \begin{minipage}[t]{0.49\linewidth}
        \centering
        \includegraphics[height=3.15cm]{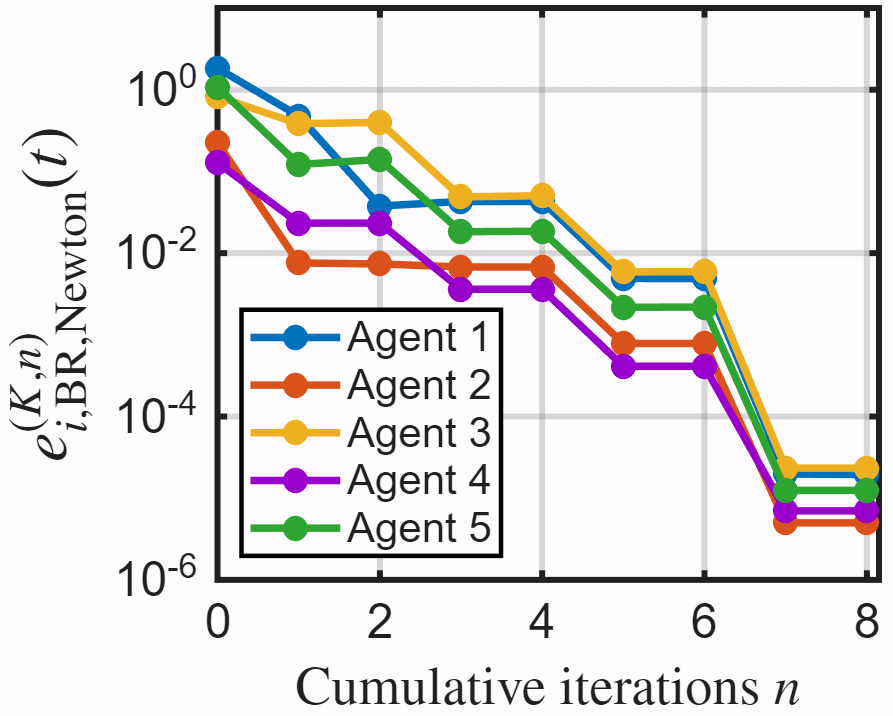}\\[-1pt]
        {\footnotesize (a)}
        \label{fig:brinstnewton}
    \end{minipage}
    \hfill
        \begin{minipage}[t]{0.49\linewidth}
        \centering
        \includegraphics[height=3.15cm]{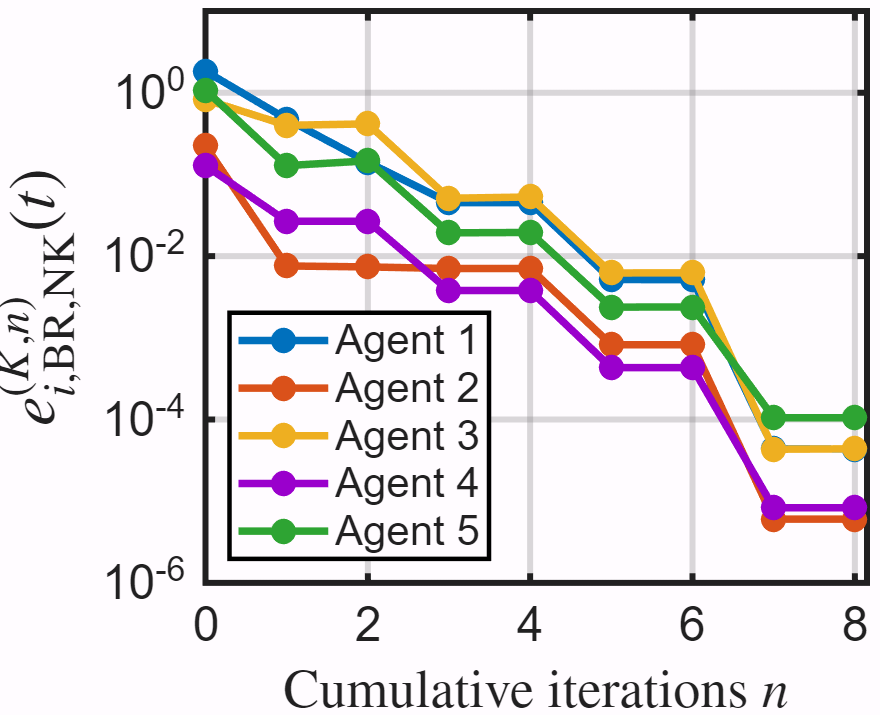}\\[-1pt]
        {\footnotesize (b)}
    \end{minipage}
    \caption{\footnotesize Error over iterations at $t=13$ and $K=2$ using better response algorithm with: (a)  Newton method and (b) Newton--Kantorovich method.}
    \label{fig:br_inst}
\end{figure}

For specific $t$ (e.g., $t = 13$), Fig.~\ref{fig:br_inst} shows error $e_{i, \mathrm{BR},M}^{(K,n)}(t)$ for each agent $i$ using the Newton and Newton–Kantorovich methods, respectively, where $n$ is the number of cumulative iterations. The errors are evaluated relative to $\mathbf{u}_{i,\mathrm{BR}}^{\mathrm{GS}}(t)$. For both methods, the errors decrease overall as the iterations proceed.

% \begin{figure}
%     \centering
%     \includegraphics[width=0.5\linewidth]{brtime.png}
%     \caption{Enter Caption}
%     \label{fig:placeholder}
% \end{figure}
\begin{figure}[h]
    \centering
    \includegraphics[width=0.6\linewidth]{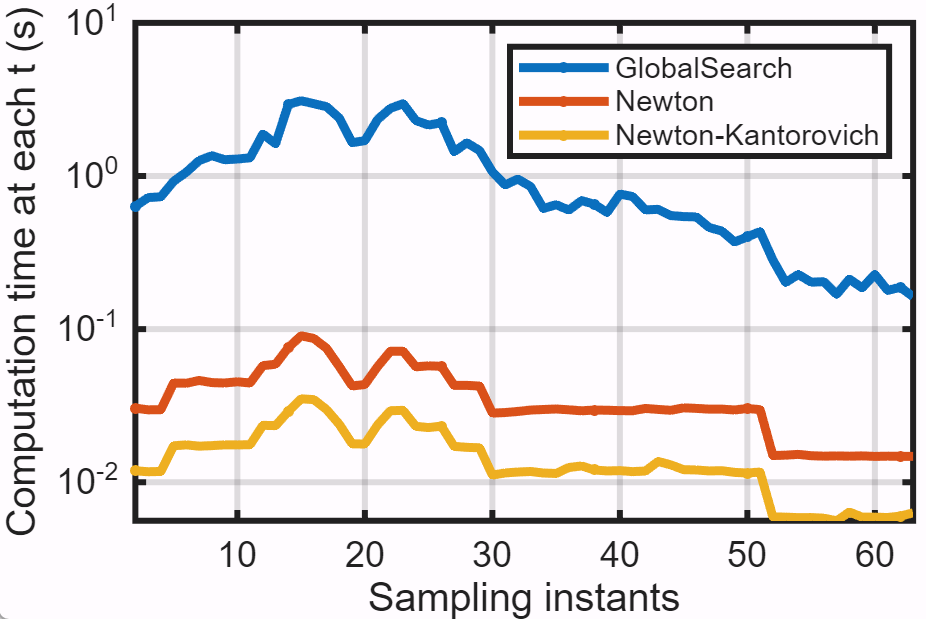}
    \caption{\footnotesize Computation time of GlobalSearch, Newton, and Newton--Kantorovich in better response algorithm.}
    \label{fig:br_solve_time}
\end{figure}

Fig.~\ref{fig:br_solve_time} compares the computation time of the GlobalSearch, Newton and Newton--Kantorovich methods in the better response algorithm. For Newton and Newton--Kantorovich methods, we fix $K=3$. As shown in Fig.~\ref{fig:br_solve_time}, Newton method significantly reduces the computation time compared to GlobalSearch method. Newton--Kantorovich method is even faster than Newton method.

Compared with the potential function optimization algorithm, the better response dynamics algorithm requires substantially more computation time when GlobalSearch is used throughout the simulation. This is because each better response sweep solves a separate GlobalSearch optimization problem for every player, whereas the potential function optimization algorithm solves only one joint optimization problem at each sampling instant. For the Newton and Newton–Kantorovich methods, the better response dynamics algorithm is faster near the beginning and the end of the simulation because fewer better response sweeps are required. Around sampling instant 20, more better response sweeps are required, causing the computation time to increase.

%Compared with the potential function optimization algorithm, the better response dynamics algorithm requires more computation time to find the NE in all three methods. This is because the better response dynamics require solving multiple optimization problems to find the NE, while the potential function optimization only needs to solve one optimization. 

\section{Conclusion}

This paper develops time-distributed Newton and Newton--Kantorovich methods to solve the GT-MPC problems in autonomous driving. These methods allow real-time NE-seeking by distributing the solution-seeking iterations over time and by performing only a fixed number of iterations at each $t$ to approximate the NE. Numerical results in a 5-vehicle intersection-crossing scenario show that these two methods can effectively facilitate both the potential game optimization and the better response dynamics algorithms, respectively, to reduce the computation time, and that the approximation errors remain bounded in all algorithms/methods during the simulated trajectories. Results also suggest that compared to the Newton methods, the Newton--Kantorovich method uses less computation time to fulfill one iteration, while it may need more iterations at each $t$ to achieve similar approximation performance as the Newton method.

%update approximate solutions online with limited computation at each sampling instant, making them suitable for real-time multi-agent decision-making. Numerical experiments with both the potential function and better response algorithms show that the Newton and Newton-Kantorovich methods improved computational efficiency while maintaining good approximation accuracy. In both algorithms, increasing the number of iterations improves the approximation accuracy, while both the Newton-type methods require much less computation time than the GlobalSearch method. Moreover, the better response algorithm shows smaller control sequence errors and a smaller gap between the Newton and Newton--Kantorovich methods, indicating that it is more favorable for the fixed-Hessian approximation. These results show that time-distributed Newton-type methods provide a promising way for real-time GT-MPC in autonomous driving.

\bibliography{ifacconf}
           % bib file to produce the bibliography
                                                     % with bibtex (preferred)
                                                   
%\begin{thebibliography}{xx}  % you can also add the bibliography by hand

%\bibitem[Able(1956)]{Abl:56}
%B.C. Able.
%\newblock Nucleic acid content of microscope.
%\newblock \emph{Nature}, 135:\penalty0 7--9, 1956.

%\bibitem[Able et~al.(1954)Able, Tagg, and Rush]{AbTaRu:54}
%B.C. Able, R.A. Tagg, and M.~Rush.
%\newblock Enzyme-catalyzed cellular transanimations.
%\newblock In A.F. Round, editor, \emph{Advances in Enzymology}, volume~2, pages
%  125--247. Academic Press, New York, 3rd edition, 1954.

%\bibitem[Keohane(1958)]{Keo:58}
%R.~Keohane.
%\newblock \emph{Power and Interdependence: World Politics in Transitions}.
%\newblock Little, Brown \& Co., Boston, 1958.

%\bibitem[Powers(1985)]{Pow:85}
%T.~Powers.
%\newblock Is there a way out?
%\newblock \emph{Harpers}, pages 35--47, June 1985.

%\bibitem[Soukhanov(1992)]{Heritage:92}
%A.~H. Soukhanov, editor.
%\newblock \emph{{The American Heritage. Dictionary of the American Language}}.
%\newblock Houghton Mifflin Company, 1992.

%\end{thebibliography}

                                                                         % in the appendices.
\end{document}